# Perpendicular magnetization of $Co_{20}Fe_{50}Ge_{30}$ films induced by MgO interface

Manli Ding and S. Joseph Poon

*Department of physics, University of Virginia, Charlottesville, Virginia 22904, USA*

Epitaxial growth of $Co_{20}Fe_{50}Ge_{30}$ thin film on single crystal MgO (001) substrate is reported. Structure characterization revealed (001)-oriented B2 order of CoFeGe well lattice matched with the MgO barrier. Perpendicular magnetic anisotropy (PMA) was achieved in the MgO/CoFeGe/MgO structure with an optimized magnetic anisotropy energy density ($K$) of $3 \times 10^6$ erg/cm$^3$. The magnetic anisotropy is found to depend strongly on the thickness of the MgO and CoFeGe layers, indicating that the PMA of CoFeGe is contributed by the interfacial anisotropy between CoFeGe and MgO. With reported low damping constant, CoFeGe films are promising spintronic materials for achieving low switching current.

Ferromagnetic materials possessing perpendicular magnetic anisotropy (PMA) are of great research interest as they have a potential for realizing next-generation high-density spin transfer torque random access memory. For applications in spintronic devices, materials possessing high thermal stability at low dimension and low current-induced magnetization switching are highly desirable.[1-2] However, the conventional PMA materials, including Co/(Pd, Pt) multilayers,[3–4] $L1_0$-ordered (Co,Fe)Pt alloy[5] or rare earth/transition metal alloys[6] are not successful to satisfy these requirements. Recently, strong interfacial perpendicular anisotropy between CoFeB and MgO, especially in Fe-rich composition,[7] was shown with a high tunnel magnetoresistance ratio, over 120%, high thermal stability at dimension as low as 40nm diameter and a low switching current of 49 μA.[8] However, the damping constant (α) increases sharply with the decreasing thickness of magnetic film, resulting in the difficulty of the reduction in the intrinsic critical current density ($J_{c0}$) of low current-induced magnetization switching. PMA alloy films with high spin polarization and low damping constant are promising spintronic materials for the further reduction in $J_{c0}$.[9] It has been recently demonstrated that $(CoFe)_{100-x}Ge_x$ films possess B2 order and very low damping constant of α ~0.0025 in the composition range of 20 at. %≤$x$≤30 at. % Ge after annealing,[10] which is significantly lower than the 0.01 of CoFeB.[11] In this letter, we fabricated Fe-rich $Co_{20}Fe_{50}Ge_{30}$ (CoFeGe) thin films faced to MgO layer and demonstrated the perpendicular magnetization of CoFeGe thin films.

CoFeGe films in this study were prepared on single crystalline MgO (001) substrates by RF magnetron sputtering system. The base pressure of the sputtering chamber was below $7 \times 10^{-7}$ Torr. Cubic MgO (001) substrates were selected to provide a seeding effect. Before the deposition, all the MgO substrates were annealed at 150 °C for 30 min *in situ* to remove the surface contamination such as $H_2O$ for a better seeding condition. Subsequently, the CoFeGe alloy films were deposited by means of co-sputtering with the elemental targets under a processing pressure around $6 \times 10^{-3}$ Torr. The capping MgO layer was formed directly from a sintered MgO target to protect CoFeGe layer from oxidation. All the samples deposited at room temperature have typical structure consisting of MgO(001)/CoFeGe($t$)/MgO(5 nm) ($t$ is the thickness of CoFeGe layer ). Subsequently, the stacks were annealed at a temperature $T_a$ of 250 °C in a vacuum under a perpendicular magnetic field of 1400 Oe for two hours. The layer thickness was identified by an x-ray reflectometer. The composition of CoFeGe film was determined using inductively coupled plasma-mass spectrometry (ICP-MS) after chemically dissolving the films, and confirmed by X-ray fluorescence (XRF) using Co, Fe and Ge standards. Structural characterization of the films were performed by x-ray diffraction (XRD) with Cu $K\alpha$ ($\lambda$ = 1.541 Å) radiation (Smart-lab®, Rigaku Inc.) and high resolution transmission electron microscopy (HRTEM). All measurements were performed at room temperature.

To investigate the structure of CoFeGe films, x-ray spectrum ($\theta$-$2\theta$ profile) is shown in Fig. 1 (a), which was obtained from the stack with CoFeGe(20 nm) after annealing. The strongest peak at about 42.9° is due to the diffraction of (002) MgO planes. Two small diffraction peaks were observed around 31.17° and 65.01° in the $2\theta$ scan, which correspond to (002) and (004) CoFeGe planes. This indicates that the fabricated CoFeGe film has a (001)-oriented B2 order, characterized by total disorder between Fe and Ge, while Co atoms occupy regular site.[10] The in-plane epitaxial relationship between the CoFeGe film and MgO was determined by the $\phi$ scan of the reflection from the (022) planes of CoFeGe and MgO. The results are displayed in Fig. 1(b). The four CoFeGe (022) peaks are due to the four-fold symmetry along the c-axis of CoFeGe, which also applied to the MgO (022) peaks. A separation of 45° from the positions of (022) peaks for the CoFeGe film to those of (022) peaks for the MgO substrate, confirms their epitaxial relationship to be CoFeGe(001)[100]//MgO(001)[110]. Using Bragg's law, the lattice constant a ~ 5.736 Å can be determined, which is nearly equal to the reported value.[12] The small misfit (−3.8%) between CoFeGe and MgO in this orientation leads to highly crystalline coherent, and chemically stable epitaxial MTJ structures. To investigate the microstructure of CoFeGe film, cross-sectional HRTEM images was taken from the annealed stack with CoFeGe(2nm). To identify the phases, we measured the $d$ spacing from the fast Fourier transformation (FFT) of the HRTEM image. We cannot unambiguously identify the (002) texture. However, CoFeGe (110) planes are evident, indicating a body centered cubic (bcc) symmetry in thin films.

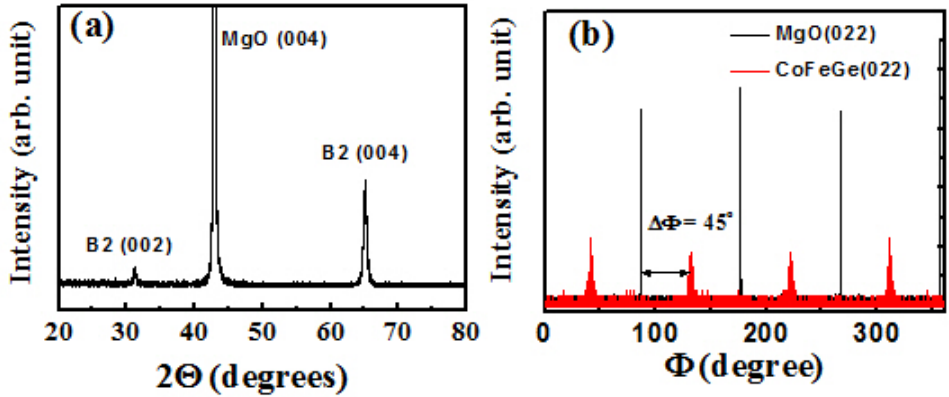

FIG. 1. (Color online) (a) XRD $\theta$-$2\theta$ scan of $Co_{20}Fe_{50}Ge_{30}$ (10 nm) film on MgO(001). (b) $\Phi$ scan on the $Co_{20}Fe_{50}Ge_{30}$ (10 nm) sample. The black line represents data taken at $2\theta$ of 44.66° (CoFeGe (022) peak) and the red one represents data taken at 62.45° (MgO (022) peak).

The magnetic behaviors were characterized along MgO [100] and [001] directions using the VSM option in Quantum Design VersaLab. The interfacial perpendicular anisotropy in thin CoFeGe films appeared after annealing, whereas magnetic easy axis was in-plane before annealing for all thicknesses, which may be due to the improvement of crystallization of the CoFeGe layer with optimized interface. Figure 2 shows the typical in-plane and out-of-plane hysteresis loops of the magnetization for the annealed stacks of MgO(001)/CoFeGe(*t*)/MgO(5 nm) with nominal thickness *t*=1.2 nm and 2 nm, respectively. The sample with $t_{CoFeGe} = 2$ nm has an easy in-plane axis. A clear perpendicular anisotropy was realized with the in-plane saturation field $H_k$=5500 Oe and out-of-plane coercivity $H_c$=10 Oe for *t*=1.2 nm. The saturation magnetization is 903 emu/cc. The total perpendicular anisotropy energy density *K* at this thickness, which determines the thermal stability, is $2\times 10^6$ erg/cm$^3$, as calculated by evaluating the area enclosed between the in-plane and perpendicular M-H curves.[13] This method was acceptable for a semi-quantitative comparison given the isothermal measurement conditions, at which the sample was near the equilibrium state. The squareness of out-of-plane hysteresis loop, however, is not perfect.

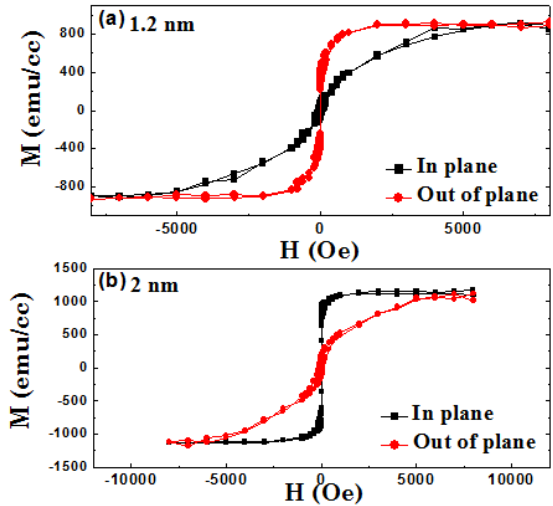

FIG. 2. (Color online) In-plane and out-of-plane magnetic hysteresis loops for (a) 1.2 nm and (b) 2 nm annealed $Co_{20}Fe_{50}Ge_{30}$ films, respectively.

It was also noticed that the magnetization $M_s$ of the film was reduced as the thickness was reduced (Fig. 3 circles). Such a trend implied the existence of a dead layer, i.e., the deteriorated crystalline structure or an intermixing between CoFeGe and the MgO layer. To determine the thickness of the dead layer and extract the saturation magnetization $M_s$ for CoFeGe in this series of samples, we linearly fitted the magnetization per unit area, $m_s=M_s t_{CoFeGe}$, as the film thickness ranged from 1.0 nm to 2.5 nm (Fig. 3 triangles). The x-intercept is the dead layer thickness, which was found to be 0.4 nm, while the slope gives the extracted $M_s$ of 1150 emu/cc. This value is consistent with the experimental bulk magnetization.[12] It is not clear whether the presence of such a dead layer has any impact on the magnetic anisotropy. More film microstructure studies are undergoing to determine the origin of the "dead layer".

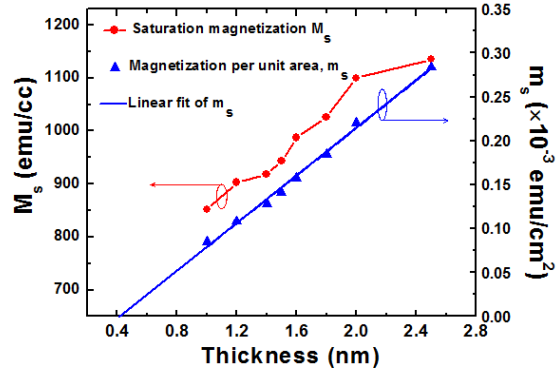

FIG. 3. (Color online) Saturation magnetization $M_s$ (emu/cc) and magnetization per unit area $m_s$ (emu/cm$^2$) vs the CoFeGe film thickness at 300 K.

To separate the bulk and interfacial contribution of the anisotropy, the overall magnetic anisotropy energy density $K$, obtained from magnetization measurements, can be expressed by the equation $K=K_b-2\pi M_s^2 +K_i/t_{CoFeGe}$ [see solid line in Fig. 4], where $K_b$ and $K_i$ are the bulk and interfacial anisotropy energy densities, respectively, and $2\pi M_s^2$ is the demagnetization energy density. From the plot, it is clearly the anisotropy changes from out-of-plane to in-plane direction with the increase in CoFeGe film thickness. This transition occurs at thickness around 1.4 nm, indicating that the interface plays an important role for PMA. From the intercept to the vertical axis, $K_i$ is estimated to be 0.9 erg/cm$^2$, which is comparable to that of CoFeB/MgO.[8] $K_b$ is obtained to be $-1 \times 10^5$ erg/cm$^3$ from the slope of the linear extrapolation, indicating the bulk anisotropy contributes to the in-plane anisotropy. Thus, the perpendicular anisotropy in this structure is entirely due to the interfacial perpendicular anisotropy between MgO and CoFeGe.

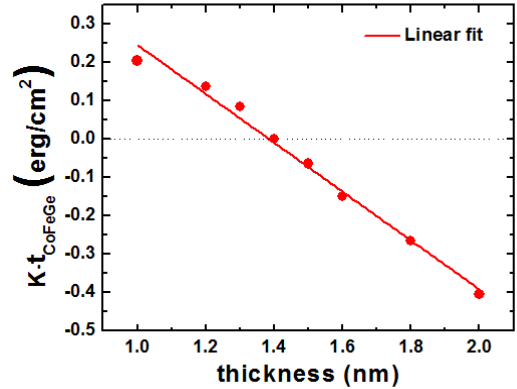

FIG. 4. (Color online)The dependence of total perpendicular anisotropy $K$ on film thickness, plotted as the product of K and thickness vs. thickness.

The interfacial perpendicular anisotropy between oxide and ferromagnetic metal has been predicted by first-principles calculation and attributed to hybridization of Co or Fe 3$d$ and O 2$p$ orbitals[14] at the interface of MgO and CoFeGe. The hybridization of 3$dz^2$ and 2$pz$ orbitals reduces the binding energy of Co–O or Fe–O perpendicular to the interface between CoFeGe and MgO, which becomes lower than that of orbitals lying in the plane, resulting in strong PMA at the interface.[14,15]

The PMA originates at the interface between the CoFeGe and MgO and is very sensitive to the thickness of each layer. Based on above, the effect of the thickness of CoFeGe ($t_{CoFeGe}$) is studied using samples with fixed thickness of MgO layer ($t_{MgO}$) (5.0 nm). The effects of $t_{MgO}$ on the magnetic properties of the annealed MgO(001)/CoFeB/MgO multilayer films are, therefore, examined here. With the CoFeGe layer thickness kept at an optimized 1.0nm, In-plane and out-of-plane hysteresis loops were taken from MgO(001)/CoFeB (1.0 nm)/MgO($t_{MgO}$)/Ta(5nm). The Ta capping here was to protect films from oxidation. The samples with $t_{MgO}$ = 0 nm exhibit in-plane anisotropy with a high anisotropic field of about 1 T (not shown). Notably, this sample has a considerable residual in-plane magnetic component, which has not been observed in previous studies. PMA emerges with the insertion of 0.8-nm-thick MgO layer, revealing that the interface between MgO and CoFeGe is essential for PMA. PMA increases with increasing $t_{MgO}$ and is greatest in the sample with $t_{MgO}$ = 2.0 nm, as shown in Fig. 5. The PMA energy density $K$ at $t_{MgO}$ = 2.0 nm can be calculated to be $3 \times 10^6$ erg/cm$^3$, which is comparable to that of other PMA systems such as CoFeB/MgO[16] and Co/Pd perpendicular multilayers,[17] and can satisfy the high thermal stability for devices with lower than 50 nm dimension.

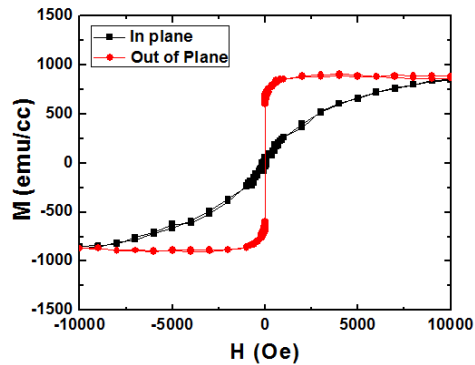

FIG. 5. (Color) In-plane and out-of-plane hysteresis loops for the annealed sample of MgO(001)/CoFeB (1.0 nm)/MgO (2.0nm)/Ta(5nm).

In conclusion, we have fabricated (001)-oriented $Co_{20}Fe_{50}Ge_{30}$ epitaxial films with the $B2$ crystal structure. We have demonstrated the perpendicular easy axis in Fe-rich CoFeGe thin film originating from the interfacial perpendicular magnetic anisotropy between CoFeGe and MgO. Its perpendicular magnetic anisotropy is found to depend strongly on the thickness of the CoFeGe and MgO. A perpendicular magnetic anisotropy energy density of $3 \times 10^6$ erg/cm$^3$ was achieved in the CoFeGe thin films, which suggests the feasibility of the application of CoFeGe to the perpendicular ferromagnetic electrodes of MTJs with high thermal stability and low magnetic damping at reduced dimension.

ACKNOWLEDGMENTS
   The authors gratefully acknowledge financial support provided by Grandis Inc. through DARPA (Award No.HR0011-09-C-0023). They thank Xiaopu Li for technical assistance.

a)corresponding author: md3jx@virginia.edu, sjp9x@virginia.edu.